\documentclass[journal]{IEEEtran}

\usepackage{cite}

\ifCLASSINFOpdf
  \usepackage[pdftex]{graphicx}
\else
  \usepackage[dvips]{graphicx}
\fi

\usepackage[cmex10]{amsmath}
\usepackage{dblfloatfix}

\usepackage{esdiff} % use this package for derivative and partial derivate commands of \diff{}{} and \diffp{}{}

\usepackage{hhline}

\begin{document}

\title{Optical Hardware Accelerators using Nonlinear Dispersion Modes for Energy Efficient Computing}

\author{Bahram~Jalali,~\IEEEmembership{Fellow,~IEEE,}
        and~Ata~Mahjoubfar,~\IEEEmembership{Member,~IEEE}% <-this % stops a space
\thanks{B. Jalali and A. Mahjoubfar are with the Electrical Engineering Department, University of California, Los Angeles, CA 90095, USA and California NanoSystems Institute, Los Angeles, CA 90095, USA.}% <-this % stops a space
\thanks{B. Jalali is with the Department of Bioengineering, University of California, Los Angeles, CA 90095.}% <-this % stops a space
\thanks{Manuscript received September 30, 2014.}}

% The paper headers
\markboth{Submitted to Proceedings of the IEEE on September 30, 2014, DOI: 10.1109/JPROC.2015.2418538}%
{Jalali \MakeLowercase{and} Mahjoubfar: Nonlinear Dispersion Modes}

% make the title area
\maketitle

\begin{abstract}
This paper proposes a new class of hardware accelerators to alleviate bottlenecks in the acquisition, analytics, storage and computation of information carried by wideband streaming signals.
\end{abstract}


\IEEEpeerreviewmaketitle

\section{Introduction}

\IEEEPARstart{H}{ardware} accelerators are custom digital computing engines with successful examples including communication and media processors \cite{owens2007survey,manavski2007cuda,porrmann2002implementation}. As opposed to general purpose computers that perform a myriad of applications specified by the software, hardware accelerators are optimized to perform a specific function but do so faster and with less power than a general purpose processor. Here we introduce the concept of photonic hardware accelerators (PHA) -- analog real-time optical waveform transformations that alleviate bandwidth and sensitivity bottlenecks in acquisition and storage of wideband streaming data as well as perform real-time analytics and coding.  In the context of computing, PHA can be viewed as the optical rendition of analog computers that appeared in the early days of computing. While analog computers were replaced by much more powerful and versatile digital computers, the lack of power efficient and scalable logic operations in optics makes it unlikely that analog optical computing will be replaced by a digital optical counterpart.

\begin{figure}[!t]
\centering
\includegraphics[width=3.4in]{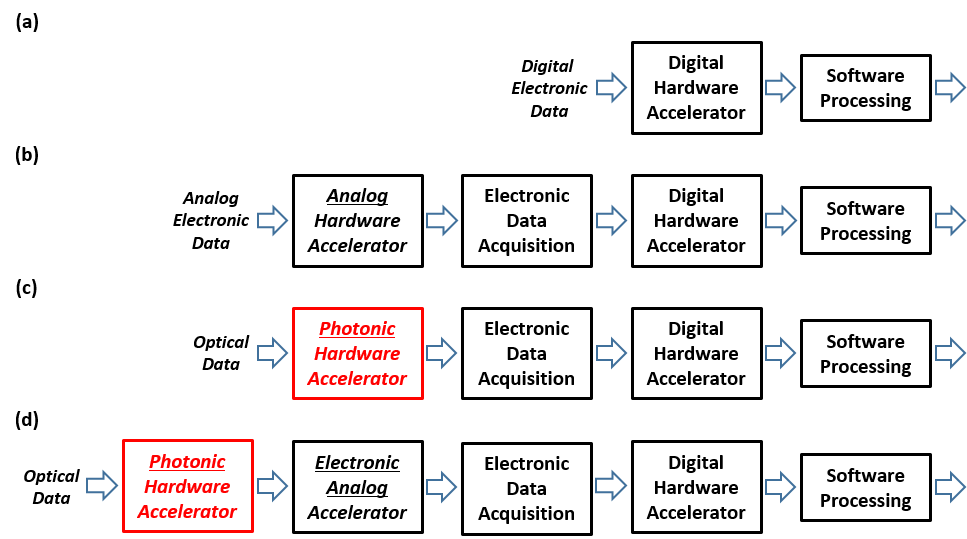}
\caption{Comparison of traditional hardware accelerator with analog and photonic hardware accelerators (PHA). (a) Traditional hardware accelerators are custom digital signal processing devices designed to perform computational operations that would be slow and power hungry if performed in software. (b) An analog hardware accelerator can be placed before the analog-to-digital conversion to take part of the processing burden off of the digital hardware and software processing. (c) Photonic hardware accelerators are analog computing engines that make it possible to capture and digitally process wideband optical data in real-time. (d) Combinations of photonic hardware accelerators and electronic analog accelerators as analog coprocessors before the analog-to-digital conversion are even more efficient in supplementing the functions of the digital processors.}
\label{fig:Hardware_Accelerator}
\end{figure}

Traditional hardware accelerators (\figurename~\ref{fig:Hardware_Accelerator}) are application-specific digital signal processing devices designed to perform computational operations, such as fast Fourier transform (FFT) and video encoding, that would be slow and power hungry if performed in software. By doing so, they increase the speed and reduce the power dissipation. In contrast, PHAs are analog optical engines. One type of PHA reduces the envelope bandwidth of fast optical data to allow digitization by a much lower bandwidth electronic Analog-to-Digital Converter (A/D converter or ADC). In doing so, it makes it possible to capture and digitally process wideband optical data in real-time. Also, by taking advantage of the sparsity in the signal it reduces the amount of data produced in wideband A/D conversion. Such an PHA simultaneously addresses two of the main bottlenecks in high speed real-time data acquisition, i.e. the bandwidth requirement for capturing ultra wideband signals and subsequently, storage and management of the large amount of data produced by wideband A/D. In one implementation it offers data compression, but in an entirely different fashion than compressive sensing \cite{bosworth2013high,valley2012compressive} and one that is far better suited for real-time operation.

\begin{figure}[!t]
\centering
\includegraphics[width=2.5in]{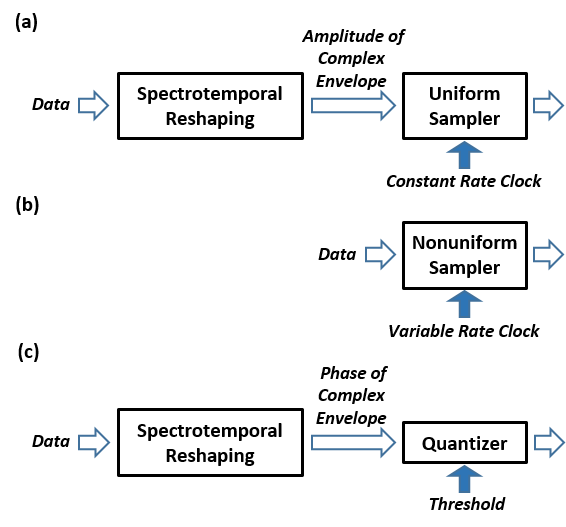}
\caption{Conventional nonuniform sampling vs. the new method for achieving the same via spectrotemporal warping followed by uniform sampling. (a) The central idea is a reshaping, specifically warping of the signal such that after uniform sampling, more samples are allocated in the information-rich (high-entropy) portion than the sparse regions. (b) While the concept of nonuniform sampling is not new, what is new is achieving this through reshaping the signal followed by uniform sampling, as opposed to using a variable-rate sampler. (c) The same class of transformations provide a phase output that is sensitive to transitions in the data and combined with thresholding, offer a mean to detect fast rapid changes and anomalies (see Section \ref{scn:digital_implementation}).}
\label{fig:Nonuniform_Sampling}
\end{figure}

\figurename~\ref{fig:Nonuniform_Sampling} depicts the concept of the information engineering leading to nonuniform sampling of envelope amplitude. Both are achieved by transformation of the signal, prior to sampling, using group delay dispersion primitives described in the main body of this paper. In \figurename~\ref{fig:Nonuniform_Sampling}a, the central idea is a reshaping, i.e. a warping of the signal that, upon subsequent uniform sampling, causes more samples to be allocated to the information-rich portion of the signal than the sparse regions. While the concept of nonuniform sampling is not new, what is new here is achieving this through reshaping the signal followed by uniform sampling, as opposed to using a variable-rate sampler (\figurename~\ref{fig:Nonuniform_Sampling}b). This approach obviates the need for having a dynamically tunable sampler. The same class of transformations provides a phase output that is sensitive to transitions in the data and offers a means to detect fast rapid changes and anomalies (\figurename~\ref{fig:Nonuniform_Sampling}c). Analog optical analytics is another fortuitous property of proposed dispersive reshaping operations.

\begin{figure*}[!t]
\centering
\includegraphics[width=6in]{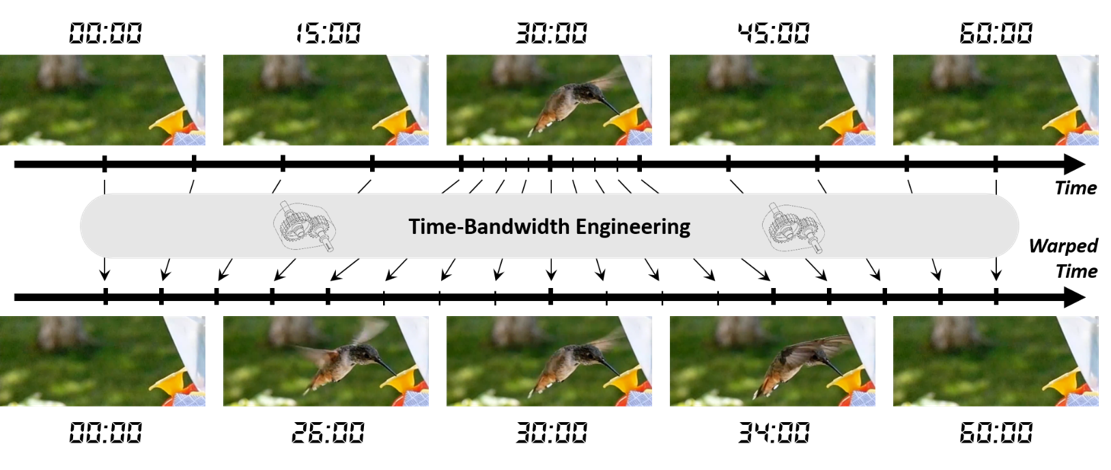}
\caption{Metaphorical visualization of the benefit offered by nonuniform sampling that exploits sparsity in the data. A video consisting of frames uniformly distributed in time lacks sufficient frame rate to capture the key event while produces redundant frames in the sparse portions. In the proposed systems, nonuniform sampling is achieved by warping the signal prior to uniform sampling.}
\label{fig:Bird}
\end{figure*}

The analog processing engine contemplated here may be interpreted as an optical information gearbox that matches the information rate of the signal to that of the data capture, processing, and storage blocks. The benefit can be metaphorically visualized in \figurename~\ref{fig:Bird}. The figure illustrates the analogy to a sparse video where key frames are clustered in time. A video consisting of frames uniformly distributed in time results in inefficient data representation. It lacks sufficient frame rate to capture the key events while producing redundant frames in the sparse portions. In digital video compression (such as MPEG), this problem is solved by interframe compression, where for most frames, the encoder only saves the change with respect to the prior frame \cite{le1991mpeg,richardson2004h}. The challenge and one that we address here is how to perform similar compression on analog signals in real-time and at the speed of light.

The transformations discussed in this paper, exploit signal sparsity to achieve: (1) warped Fourier domain compression, (2) enhance the analog signal-to-noise ratio during data acquisition, and (3) perform analytics. The operations are performed using filters with large and reconfigurable group delay dispersion. Such filters are most readily realizable in optics because of availability of wideband, low loss, and reconfigurable dispersive elements. In discussions that follow we assume information is already in the optical domain, or has been modulated onto an optical carrier. We will not be concerned with how the data is generated and about the electro-optic conversion process. While the focus of the present paper is on analog signals, the processes can be emulated in discrete domains and numerically applied to digital data as discussed in Section \ref{scn:digital_implementation}. These represent a new class of digital signal processing algorithms that are inspired by the physics of dispersive propagation of temporal signals, and diffractive propagation of images, through media with specifically designed curved (nonlinear) dispersion and diffraction properties.

\section{Mathematical Framework}

Let's consider operating on a signal with a phase filter described by the transfer function
\begin{equation}
\label{eqn:transfer_function}
H(\omega) = \exp\lbrack i \phi(\omega) \rbrack
\end{equation}
where $\phi(\omega)$ is an arbitrary function of frequency, $\omega$, and its range can be significantly larger than $2\pi$. The phase can be expanded in terms of basis functions (or modes), $\phi_m(\omega)$:
\begin{equation}
\phi(\omega) = \sum_{m=0}^{+\infty}\phi_m(\omega)
\end{equation}
The transfer function is a phase operator, which can be viewed as a cascade of mode operators, $H_m(\omega)$:
\begin{equation}
\label{eqn:operator}
H(\omega) = \prod_{m=0}^{+\infty} H_m(\omega) = \prod_{m=0}^{+\infty}\exp\lbrack i \phi_m(\omega) \rbrack
\end{equation}
The basis functions can be polynomials or other suitable functions as discussed later. In the case of polynomials, basis functions take the form of powers of $\omega$ and can be expressed in terms of the Taylor expansion: 
\begin{eqnarray}
\label{eqn:phase_modes}
\phi_m(\omega) = \frac{\phi^{(m)}}{m!}(\omega-\omega_c)^m\\
\phi^{(m)} = \left.\diff[m]{\phi(\omega)}{\omega}\right\vert_{\omega = \omega_c}
\end{eqnarray}
where $\phi^{(m)}$ is the $m$th derivative of phase with respect to frequency evaluated at the carrier frequency, $\omega_c$. By definition, the modulation sideband frequency is $\omega_m = \omega - \omega_c$.

Within this framework, the filter group delay, $\tau(\omega)$, is given by 
\begin{equation}
\tau(\omega) = \diff{\phi(\omega)}{\omega} = \sum_{m=1}^{+\infty}\tau_m(\omega)
\end{equation}
For the case of polynomial basis functions, the group delay modes become,	
\begin{equation}
\label{eqn:group_delay_modes}
\tau_m(\omega) = \frac{\phi^{(m)}}{(m-1)!}(\omega-\omega_c)^{m-1}
\end{equation}
Similarly, $\tau_m^{-1}(\tau)$ are the chirp modes of $m$th order. 

Upon propagation through the filter, an envelope modulated pulse with spectrum $\widetilde{E}_i(\omega-\omega_c)$ is transformed into a temporal signal with an envelope given by,
\begin{equation}
\label{eqn:Eout_integral}
E_o(t)=\frac{1}{2\pi}\int_{-\infty}^{+\infty} \widetilde{E}_i(\omega-\omega_c) H(\omega) \exp\lbrack -i (\omega-\omega_c) t\rbrack d\omega
\end{equation}
The integral can be solved using the stationary phase approximation. The approximation assumes that the input spectrum is a slowly varying function while the exponential in the integral varies rapidly. The latter is satisfied when filter group delay is relatively large leading to the far field regime of dispersion. Under this condition, the spectrum is mapped into time, with the mapping relation
\begin{equation}
\label{eqn:group_delay}
\tau(\omega) = \sum_{m=1}^{+\infty}\frac{\phi^{(m)}}{(m-1)!}(\omega-\omega_c)^{m-1}
\end{equation}
The $m=1$ term is a time shift corresponding to the latency through the filter. The $m=2$ term describes linear frequency to time mapping meaning that at any time t, only a distinct frequency contributes to the temporal envelope at that time. Hence the output temporal envelope is a replica of the optical spectrum, i.e. the system performs Fourier transformation.  Fourier domain sampling is then achieved by simply sampling the dispersed temporal envelope. A powerful feature of this type of transformation is the time-stretch causing temporal envelope to be made arbitrarily slow such that the spectrum can be digitized in real-time by an analog-to-digital converter. 

The $m=2$ mode can be identified as the classic Time-Stretch Dispersive Fourier Transform (TS-DFT), a powerful technique with early application in wideband analog-to-digital conversion \cite{coppinger1998time} and spectroscopy \cite{kelkar1999time}. Described as ``slow motion at the speed of light,'' TS-DFT has since led to the discovery of optical rogue waves \cite{solli2007optical}, world record performance in analog-to-digital conversion \cite{Ng2014demonstration}, the creation of a new imaging modality known as the time stretch camera \cite{goda2009serial,qian2009real,zhang2011serial,fard2011nomarski,wong2012optical,kalyoncu2014fast}, which has enabled detection of cancer cells in blood with record sensitivity \cite{goda2012high,mahjoubfar2013label,chen2014hyper,mahjoubfar2014label}, and a portfolio of other fast real-time instruments such as ultrafast vibrometric imagers \cite{mahjoubfar2011high,yazaki2014ultrafast,goda2012hybrid,mahjoubfar20133d}. For a review of this technique please see reference \cite{goda2013dispersive}.

Higher order terms in (\ref{eqn:group_delay}) describe nonlinear, i.e. warped, mapping between frequency and time. Such a system with high enough dispersion (called far-field range) can be interpreted as performing warped Fourier transform. As will show, to achieve nonuniform Fourier domain sampling albeit using a uniform time domain sampler, the warp profile must be chosen according to the spectrotemporal sparsity of the signal.

It is important to note that the precise one-to-one mapping of frequency to time, suggested by (\ref{eqn:group_delay}) is not exact; it is subject to the accuracy of the stationary phase approximation, i.e. the degree to which we are in the far field of dispersion. In reality, not a single but rather a finite range of frequencies survives the integral in (\ref{eqn:Eout_integral}) and appear at a given time instance, t. The range is centered at frequency described by (\ref{eqn:group_delay}) and with a width given by 
\begin{equation}
\label{eqn:Resolution}
\delta\omega_m = \sqrt{\frac{4\pi}{\sum\limits_{m=2}^{+\infty}\frac{\phi^{(m)}}{(m-2)!}(\omega-\omega_c)^{m-2}}}
\end{equation}
The parameter $\delta\omega_m$ can be interpreted as the frequency resolution, or the ambiguity, in the frequency to time mapping relation. (\ref{eqn:Resolution}) can be obtained from the first zero crossings of the real and imaginary parts of (\ref{eqn:Eout_integral}) \cite{goda2009theory}.

For the case where the group delay is linear, (i.e. when $\phi^{(m)} = 0$ for $m \geq 3$) the filter performs an inverse Fourier transform and the classic linear mapping of spectrum into time. For higher order modes, the operation is a nonuniform warped mapping in which the strength and shape of the curve can be reconfigured through the amplitude and sign of $\phi^{(m)}$ for $m \geq 3$. One application for this is warped time stretch where a broadband waveform that is short in time is stretched making real-time digitization possible. At the same time, the amplitude of the time stretched waveform is the warped spectrum of the input waveform's envelope. The shape of the warp is dictated by the group delay of the filter, which in turn is prescribed by the input signal sparsity. As shown below, time-bandwidth compression can be achieved by designing the group delay such that it is matched to the time-frequency sparsity of the input. This function can be described as warped slow motion where the degree to which the signal is slowed down matches the information content in the signal spectrum. 

\subsection{Alternative Basis Functions}

A drawback of polynomial basis functions is that to locally design the group delay function for each region of the spectrum requires a large number of modes. One can overcome this limitation by using spline functions where the spectrum is divided into regions and a different polynomial used in each region. In addition, one can employ other localized functions such as Gaussian, sigmoids, and wavelets. The basis functions can also be a set of pseudorandom dispersion modes. Such pseudorandom spectrotemporal basis functions will be useful for signals with unknown spectral sparsity pattern.

The applied group delay dispersion chirps the input waveform with a specific profile. Any chirp in the input signal will affect the outcome and must therefore be accounted in the design of the group delay. The group delay profile should correspond to the difference between the desired chirp and the chirp of the input signal.

\subsection{Properties of Phase and Group Delay Dispersion Modes}

Here we consider basis functions $\phi_m(\omega)$ in (\ref{eqn:phase_modes}) and group delay modes $\tau_m(\omega)$ in (\ref{eqn:group_delay_modes}) as constituent modes of phase and group delay functions, and their corresponding temporal and spectral transformations as dispersive stretch primitives. Modes with $m \geq 3$ describe warped group delays and hence represent warped stretch primitives. When applied to pulsed waveforms, these modes are localized in time and can be identified as spectrotemporal stretch wavelets.

\begin{figure*}[!t]
\centering
\includegraphics[width=6in]{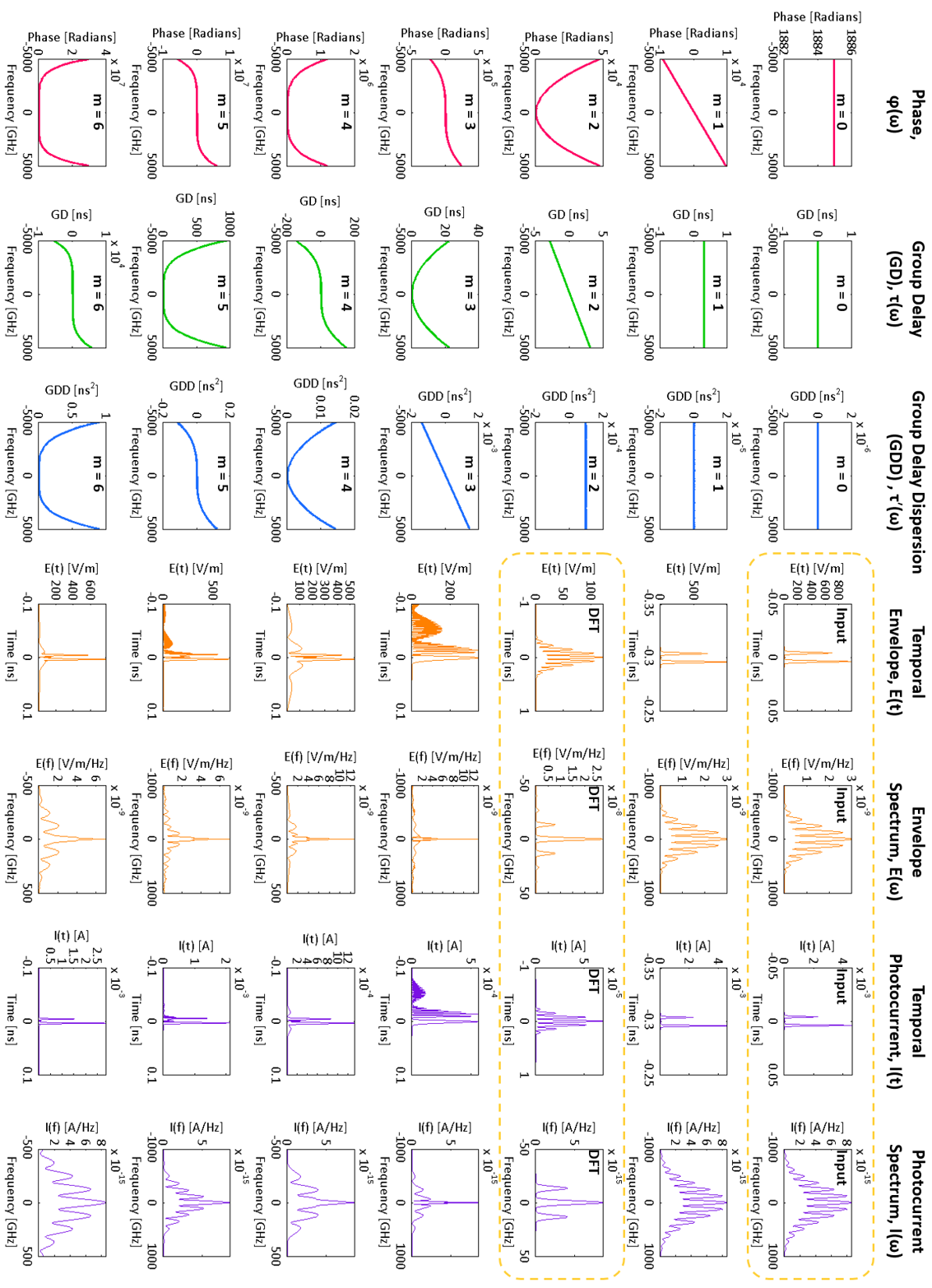}
\caption{Dispersive phase (column 1), group delay (column 2) and group delay dispersion (column 3) modes for the 7 lowest order dispersion modes. The figure also shows the stretch operations performed by these modes. Columns 4 and 5 are the time domain, $E(t)$, and spectrum, $E(\omega)$ of the complex passband signal envelope. Columns  6 and 7 are the same for the square of the magnitude, i.e. the power of the complex signal. For an optical signal this would be proportional to the photocurrent, hence column 6 and 7 are labeled as $I(t)$, and $I(\omega)$. The $m=0$ mode is a constant phase (zero group delay) which has no influence on the signal, so the output envelope is same as that of the input signal. The $m=1$ mode is the linear phase (constant group delay) operation. Its effect on the signal is simply a constant time shift (latency) with no change in the spectrum magnitude. The $m=2$ mode is the linear group delay performing the Time-Stretch Dispersive Fourier Transform (TS-DFT) operation. The $m=3$ mode as well as higher odd order modes are characterized by a group delay that is symmetric about the carrier frequency and cause folding of the spectrum during spectrum to time mapping. The $m=4$ and higher even order modes describe nonlinear mappings of spectrum into time and warped stretch transformations.}
\label{fig:modes}
\end{figure*}

\figurename~\ref{fig:modes} shows the phase mode $\phi_m(\omega)$ (column 1), group delay mode (column 2) $\tau_m(\omega)$, and group delay dispersion mode (column 3) $\tau'_m(\omega)$ for seven lowest order modes ($m=0$ to 6). The Figure also shows the transformations they perform, in subsequent columns. Columns 4 and 5 are the time domain, $E(t)$, and spectrum, $E(\omega)$, of the complex passband signal envelope. Columns 6 and 7 are the same for the modulus squared of the complex amplitude, i.e. the power of the complex signal. For an optical signal this would be proportional to the photocurrent, hence column 6 and 7 are labeled as $I(t)$, and $I(\omega)$.

The $m=0$ mode is a constant phase (zero group delay), which has no influence on the signal envelope. Hence columns 1 and 2, in the $m=0$ row, are the temporal envelope and the envelope spectrum of the input signal under test.

The $m=1$ mode is the linear phase (constant group delay) operation. Its effect on the signal is simply a constant time shift (latency) of the envelope with no change in the envelope spectrum. The $m=2$ mode is the linear group delay performing the TS-DFT operation. The spectrum is linearly mapped into time resulting in a stretched temporal waveform that is a replica of the spectrum.

The $m=3$ mode as well as higher odd order modes are characterized by a group delay that is symmetric about the carrier frequency, and a phase that is antisymmetric. Upper and lower sidebands experience the same group delay causing folding of the spectrum concurrent with spectrum to time mapping. The mapping has even symmetry, with respect to group delay (odd with respect to group delay dispersion).

The $m=4$ and higher even order modes describe nonlinear mapping of spectrum into time and warped stretch transformation. These modes exhibit preferentially higher group delay at the wings of the spectrum (away from the carrier frequency). The frequency to time mapping has odd symmetry and stretching in time is warped and occurs primarily at the wings. More complex warp structures can be synthesized by combining the $m=2$ (linear group delay) mode and higher even order modes with proper sign and amplitude as discussed below. Although phase modes linearly superimpose to form a desired group delay profile, the output envelope is not a linear superposition of the output envelopes of each mode. This is due to the nonlinear relationship between phase modes and output envelopes.

These phase and group delay operations have other utilities beyond warped Fourier domain sampling. By virtue of applying an engineered chirp to the signal, they can be used to manipulate particular regions of the frequency spectrum. Dispersion modes and stretch wavelets can be viewed as a special class of eigenmodes and wavefunctions. As shown below, the distinct symmetry of dispersion eigenfunctions determines their signal processing property. The modes can also be viewed as dispersion primitives from which complex dispersion operations can be synthesized. While polynomial functions are not universally orthogonal, we can envision other expansion, and certain classes of polynomials, that are orthogonal.

\subsection{Spectrotemporal Sparsity}

The traditional notion of sparsity, i.e. sparsity in time, is not pertinent here. Instead, sparsity (entropy) in the spectrum is the attribute that influences and guides the design of the filter's group delay profile. To show the connection between the group delay and the signal sparsity and as a design and visualization tool, we introduce the concept of spectrotemporal sparsity.

\begin{figure*}[!t]
\centering
\includegraphics[width=6in]{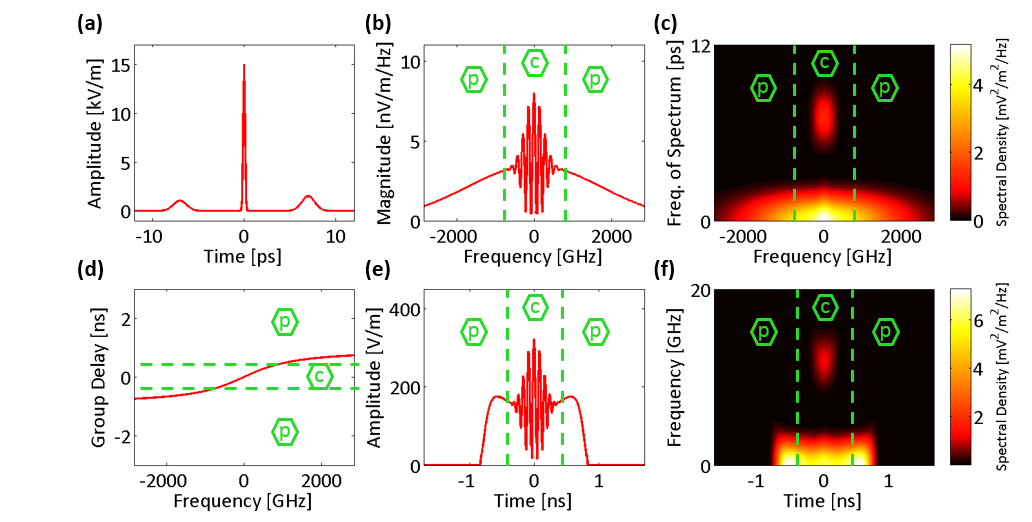}
\caption{Introducing the concept of spectrotemporal sparsity. (a) Input temporal envelope and (b) magnitude of the spectrum. (c) Local (short-time) Fourier transform of the spectrum magnitude. (d) The warped group delay dispersion used to reshape the envelope. Frequency-to-time mapping and temporal stretching are warped in such a manner that the sparse wings, i.e. the peripheral regions of the spectrum, are squeezed relative to the dense central region. (e) The sparse wings now occupy a shorter time duration leading to efficient sampling. (f) The output shows reshaped spectrotemporal distribution with energy compressed in the central region.}
\label{fig:Spectrotemporal_Simulation}
\end{figure*}

Let us consider, as an example, an optical signal whose input temporal envelope (baseband) and its spectrum are shown in \figurename~\ref{fig:Spectrotemporal_Simulation}a and \figurename~\ref{fig:Spectrotemporal_Simulation}b, respectively. \figurename~\ref{fig:Spectrotemporal_Simulation}c then shows the local (short-time) Fourier transform of the spectrum. In other words, it is equivalent to viewing the spectrum as a temporal waveform and plotting its short time Fourier transform. As a result, the horizontal axis is the input frequency and the vertical axis is the local frequency of variations in the spectrum magnitude. We call this the frequency of spectrum. In this example, input signal spectrum (\figurename~\ref{fig:Spectrotemporal_Simulation}b) is feature dense (has high entropy) in the central region and feature sparse (has low entropy) in the wings. Hence, the local frequency is high in the central region and low in the wings (\figurename~\ref{fig:Spectrotemporal_Simulation}c). While this is only one example, in the Applications section of this paper, we show a powerful optical imaging system in which the signal has this specific type of sparsity (Section \ref{scn:cell_screening}). The desired group delay must conform to the local frequency, shown in \figurename~\ref{fig:Spectrotemporal_Simulation}c. In the regions where the spectrum magnitude has fast variations, the filter (\figurename~\ref{fig:Spectrotemporal_Simulation}d) presents a high group delay dispersion (slope) resulting in larger stretching in time than the slow varying regions of the spectrum magnitude. The frequency-to-time mapping and temporal stretching are warped in such a manner that the sparse wings of the spectrum are squeezed relative to the dense central region (\figurename~\ref{fig:Spectrotemporal_Simulation}e). The low entropy wings are squeezed in so that they occupy a shorter time duration after the warped stretch. The resulting restructuring of the spectrotemporal distribution is shown in \figurename~\ref{fig:Spectrotemporal_Simulation}f.

Upon uniform sampling of the output envelope, the nonuniform mapping and warped stretch cause the dense portion of the spectrum to effectively receive higher sampling resolution than the sparse regions leading to nonuniform sampling. The sampling rate is basically matched to the signal's spectrum sparsity. We note that this nonuniform sampling is performed not by a variable rate sampler, but with a uniform sampler preceded by warped spectrotemporal reshaping. To be sure, a variable rate sampler is a device that is exceedingly difficult to realize as it would have to dynamically adapt its sampling rate to the signal in real-time.

\subsection{Complex Spectrotemporal Synthesis}

\begin{figure*}[!t]
\centering
\includegraphics[width=5in]{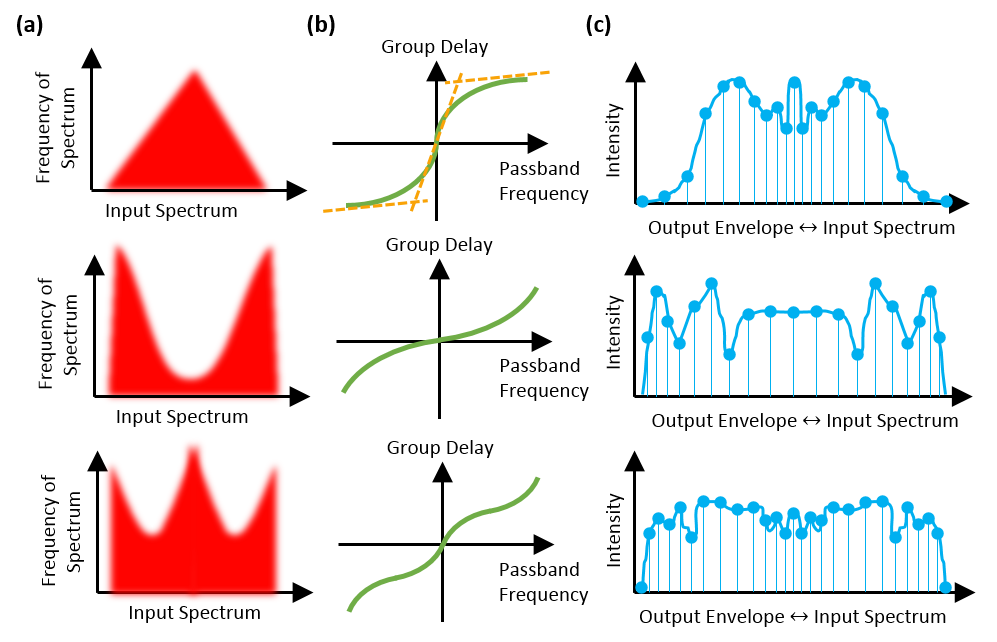}
\caption{(a) The first column shows three possible types of spectrotemporal distribution of the input signal envelope. (b) The middle column is the corresponding group delay that matches the sparsity in each case. (c) The last column shows what effectively happens when the warped signal is sampled uniformly. A lower sampling rate is effectively applied to the sparse regions and higher rate to the dense regions of the spectrum with the use of a constant rate sampler.}
\label{fig:Spectrotemporal_Drawing}
\end{figure*}

More complex group delay and stretch profiles, than those shown in \figurename~\ref{fig:Spectrotemporal_Simulation}, can be synthesized using the mode library described in \figurename~\ref{fig:modes}. Three examples are shown in \figurename~\ref{fig:Spectrotemporal_Drawing}. In each case, the group delay profile is governed by the spectrotemporal sparsity of the input waveform. \figurename~\ref{fig:Spectrotemporal_Drawing}a depicts three possible types of spectrotemporal distributions of the input signal spectrum. \figurename~\ref{fig:Spectrotemporal_Drawing}b shows the corresponding filter group delay that matches the sparsity in each case. The asymptotic dashed lines show the group delay dispersion, i.e. the slope of the group delay. In the sparse regions, the group delay dispersion is smaller resulting in a more compact mapping into time. Conversely, the dense portions of the spectrum experience a larger dispersion and are stretched into a longer time scale. Once again, upon uniform sampling and dewarping of the output envelope, the operation leads to nonuniform sampling of the input spectrum (\figurename~\ref{fig:Spectrotemporal_Drawing}c).

\section{Optical Implementations and Reconfigurability}

Such signal transformations require phase filters with large group delays. Optical fibers and waveguides provide a practical method for implementing such filters in the optical domain. While optics suffers from lack of efficient electro-optical and optical-optical interactions necessary for switching and logic, it does offer well known virtues of low loss and wideband operation. What is less appreciated, and one that is relevant to the present discussion, is its temporal dispersion property offering nanoseconds of differential group delay over tens of terahertz of bandwidth, metrics that are unimaginable in the realm of electronics.

\begin{figure*}[!t]
\centering
\includegraphics[width=6in]{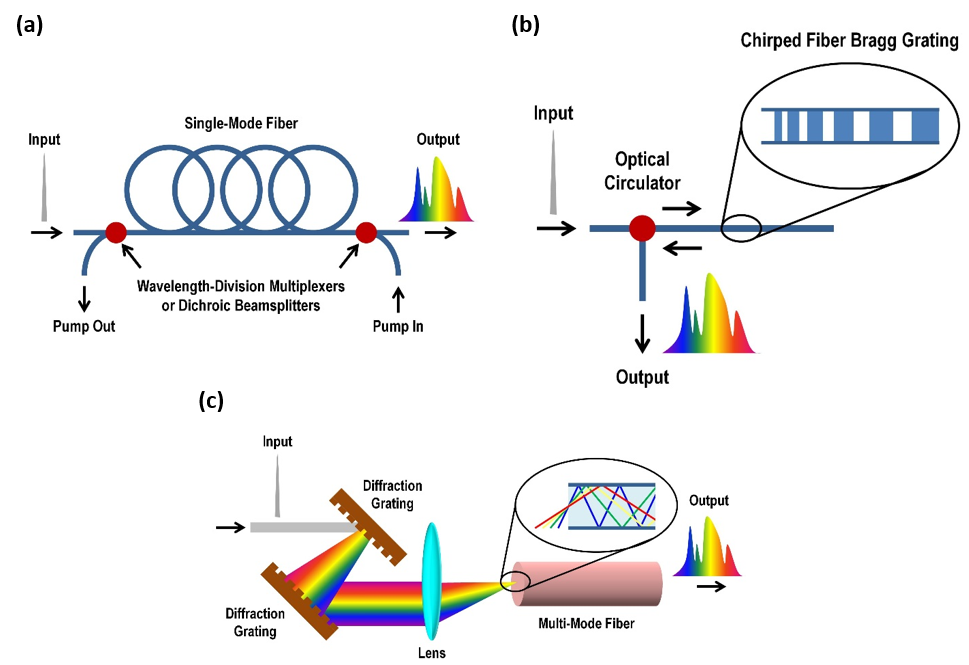}
\caption{Various method for realizing optical dispersion. (a) Dispersive optical fiber with internal Raman amplification. (b) Chirped Fiber Bragg Grating (CFBG) used in conjunction with a circulator. (c) Chromo-Modal Dispersion (CMD); The CMD device offers field reconfigurability of phase and group delay dispersion through tuning of the launch angle of light into the multi-mode waveguide.}
\label{fig:Optical_Dispersion}
\end{figure*}

Temporal dispersion in optical domain, otherwise known as group delay dispersion, describes the dependence of the propagation constant on frequency. The dependence originates from the spectrum of the material's refractive index and from that of the propagation constant of the guided mode. In frequency domain, dispersion manifests itself as a phase filter with transfer function given by the operator (\ref{eqn:operator}) and with the phase, group delay, and group delay dispersion as
\begin{eqnarray}
\phi(\omega) = \sum_{m=0}^{+\infty}\phi_m(\omega) = \sum_{m=0}^{+\infty}\frac{\beta_m}{m!}(\omega-\omega_c)^m\\
\tau(\omega) = \sum_{m=1}^{+\infty}\tau_m(\omega) = \sum_{m=1}^{+\infty}\frac{\beta_m}{(m-1)!}(\omega-\omega_c)^{m-1}\\
\tau'(\omega) = \sum_{m=2}^{+\infty}\tau'_m(\omega) = \sum_{m=2}^{+\infty}\frac{\beta_m}{(m-2)!}(\omega-\omega_c)^{m-2}
\end{eqnarray}
where $\beta_m$ is the $m$th order dispersion parameter. The most readily available and commonly used dispersive optical device is the optical fiber. Among optical fiber designs, the dispersion compensation fibers (DCFs) \cite{agrawal2007nonlinear} is particularly attractive for high dispersion and low loss. These fibers are capable of producing 100s of picoseconds of group delay dispersion per nanometer of optical spectrum (1 nm corresponds to 125 GHz at the telecommunication wavelength of 1550 nm) with a concomitant optical loss of less than 1 dB. With addition of internal amplification via the stimulated Raman scattering, the dispersion can be extended in excess of 10 ns/nm \cite{chou2007femtosecond,mahjoubfar2013optically,goda2009demonstration} (\figurename~\ref{fig:Optical_Dispersion}a). Other attractive options are the chirped fiber Bragg gratings (CFBGs) (\figurename~\ref{fig:Optical_Dispersion}b) \cite{hill1997fiber,conway2008phase}, the ChromoModal Dispersion (CMD) device (\figurename~\ref{fig:Optical_Dispersion}c) \cite{diebold2011giant}, and the reconfigurable optical add-drop multiplexers (ROADMs) \cite{klein2005reconfigurable,riza1999reconfigurable}.

The CFBG is a distributed Bragg reflector (periodic variation in the refractive index) written in a short segment of an optical fiber that reflects particular wavelengths of light and transmits all others (\figurename~\ref{fig:Optical_Dispersion}b) \cite{hill1997fiber}. The refractive index periodicity is chirped so that different wavelengths reflected from the grating undergo different time delays. When combined with a circulator as shown in \figurename~\ref{fig:Optical_Dispersion}b, the device offers large group delays of several nanoseconds per nanometer but over a smaller wavelength range than the optical fiber. The advantages over fibers are its short length and ability to customize the amount of group delay magnitude and profile at the time of device fabrication. The main disadvantage is group-velocity ripples that are converted to fast temporal modulations after frequency-to-time mapping \cite{conway2008phase}. 

\begin{figure*}[!t]
\centering
\includegraphics[width=6in]{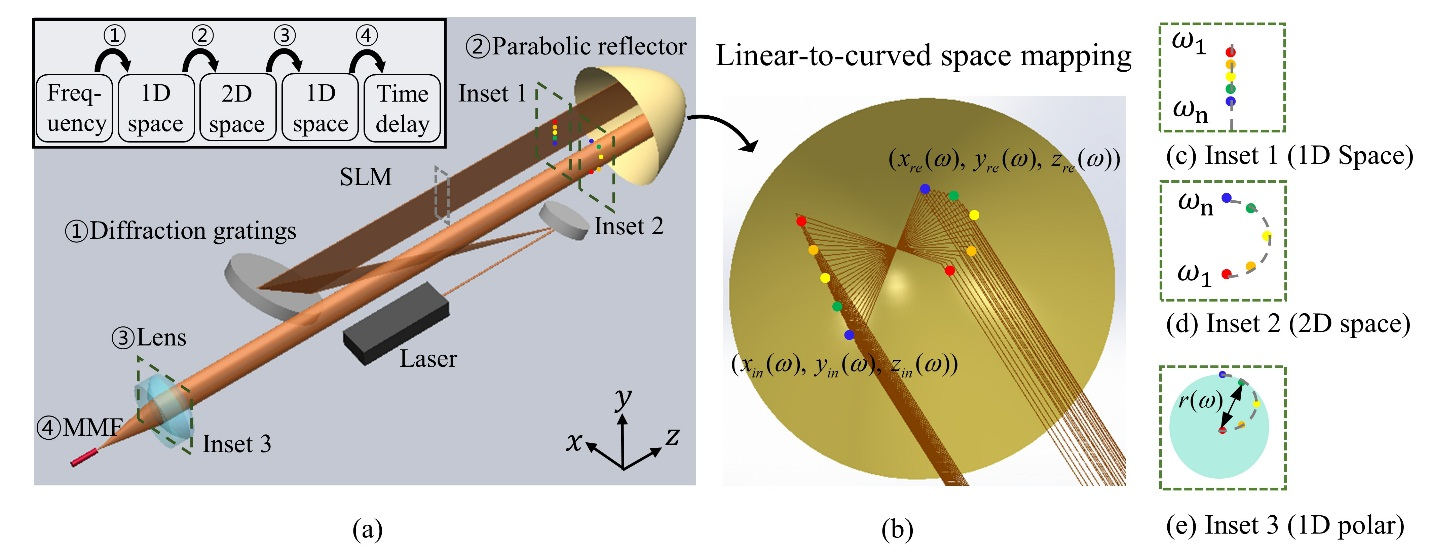}
\caption{(a) Schematic of the curved space mapping chromo-modal dispersion device (CMD device) and the functions it performs (inset). (b) Warping of spatial dispersion by a parabolic mirror. Diagrams on the right hand side show, from top to bottom, the mapping of optical frequency into one-dimensional (1D) space, two-dimensional (2D) space, and 1D polar coordinate space \cite{park2015dispersion}.}
\label{fig:curved_mirror}
\end{figure*}

The CMD  is a type of dispersive device that converts the large modal (spatial) dispersion inherent in multi-mode fibers or waveguide to chromatic (frequency dependent) dispersion (\figurename~\ref{fig:Optical_Dispersion}c) \cite{diebold2011giant}. It does so by combining the waveguide with a diffraction grating. Angularly dispersed broadband light is focused onto a multi-mode fiber coupling various spectral components into distinct fiber modes each having a different phase velocity. The strength, sign, and the delay vs. frequency curvature can be widely reconfigured by adjusting the alignment of the grating and the waveguide. The CMD device provides field reconfigurable tuning of the group delay and offers a means to achieve the type of phase filters needed for engineering the spectrotemporal structure of wideband optical waveforms in a reconfigurable fashion. Also recently, linear-to-curved space mapping is combined with the CMD to achieve a new dispersive device that offers arbitrary tuning of dispersion curvature \cite{park2015dispersion} (\figurename~\ref{fig:curved_mirror}).

The ROADM is a tunable wavelength-division multiplexing filter with a channel monitor and attenuator/amplifier that can be remotely reprogrammed to change the channel access. Having a wavelength selective switch, they can be used in parallel in conjunction with a set of tunable delays to form a quantized form of the group delay profile. As long as the spectral resolution of the channels in the ROADM is fine enough for the target application, this can be an effective approach to implement an easily reconfigurable arbitrary group delay profile.

\section{Applications}
\subsection{Nonuniform Fourier Domain Sampling: Exploiting Sparsity}

The Fast Fourier Transform (FFT) algorithm is arguably one of the most important and practical technologies in use today. By making Fourier transform computationally efficient, it has become the bedrock of digital signal processing for communication and image processing. Nevertheless, FFT has some limitations. First, real-time operation at micro- and nanosecond time intervals is still difficult for large block sizes. Second, application of FFT to optical data is restricted due to the ultra-wideband nature of optical data and the inability to digitize such broadband signals with an analog-to-digital converter. Third, FFT expands the signal into a linearly spaced basis set and this leads to a Fourier domain representation that is larger than necessary when the input signal spectrum is sparse. In fast real-time operations, this creates a big data predicament via a generation of redundant data.

The TS-DFT addresses the real-time operation and the ADC bottleneck. By mapping the spectrum into a time domain waveform that has been slowed enough to be digitized in real-time, it enables fast Fourier transformation and digitization of wideband optical signals as if it is the FFT of optics. TS-DFT has been used to create instruments that measure extremely fast optical waveforms at high throughput by measuring their Fourier spectrum as opposed to directly measuring their temporal profile.

\subsection{Case Study: Big Data Predicament in Biological Cell Screening}
\label{scn:cell_screening}

\begin{figure}[!t]
\centering
\includegraphics[width=3.4in]{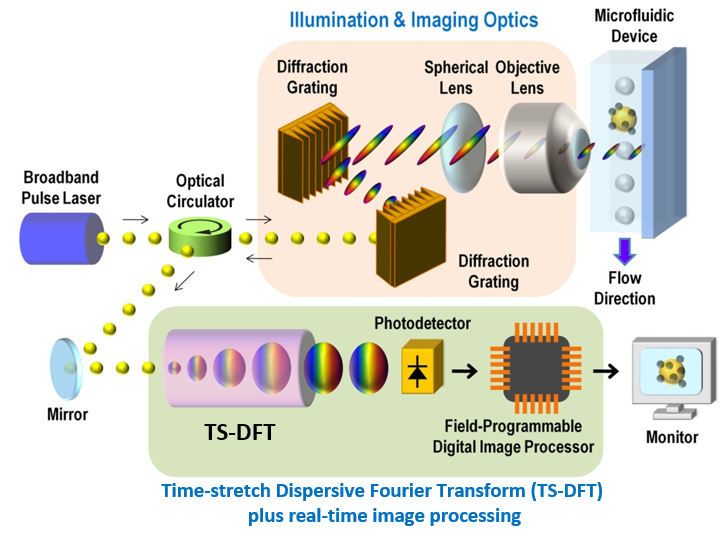}
\caption{Time stretch camera. The camera first captures fast sequential line-scans with laser pulses that are fashioned into 1D rainbows. This encodes the 1D image into the spectrum of the laser pulse. A 2D image is obtained when subsequent pulses capture line-scans along the direction of motion of cell within a microfluidic flow channel. Back-reflected pulses from the flow channel are directed via an optical circulator toward a dispersive element with linear group delay (quadratic phase) so that they can be digitized and processed in real-time. During the dispersive time stretch, images are also optically amplified to overcome the thermal noise inherent in optoelectronic conversion.}
\label{fig:STEAM}
\end{figure}

Fast real-time instruments inevitably create a big data problem.  As a case study that highlights the big data predicament in the context of imaging, we consider the time-stretch microscopy system, shown in \figurename~\ref{fig:STEAM}, for analysis of blood cells in high-speed flow. The camera takes blur-free images of fast-flowing particles in the microfluidic device. The acquired images are optoelectronically processed and screened in the real-time optoelectronic image processor that employs the TS-DFT.

The camera first captures fast sequential line-scans with laser pulses that are fashioned into 1D rainbows. This encodes the 1D image into the spectrum of the laser pulses. A 2D image is obtained when subsequent pulses capture line-scans along the direction of cell flow within a microfluidic channel. Back-reflected pulses from the microfluidic channel are directed via an optical circulator toward a dispersive element with linear group delay (quadratic phase) so that they can be digitized and processed in real-time. During the dispersive time stretch, images are also optically amplified to overcome the thermal noise inherent in optoelectronic conversion. It is also important to note that the stretched pulse duration (determined by the maximum group delay over the optical bandwidth) must be less than the laser repetition period. Hence, for a given laser repetition rate and a large bandwidth, there is an upper limit on the amount of group delay.

The time stretch imager captures up to one billion line-scans per second \cite{xing2014serial,xing20152}. Such high throughput operation is ideal for identifying rare abnormal cells in blood \cite{goda2012high,mahjoubfar2013label,chen2014hyper}. Indeed the system was successful in detecting breast cancer cells in blood with less than one-in-a-million specificity error, roughly 100 times better than the standard blood analyzer. Producing data at a rate of approximately 100 Gbps, the massive throughput of such a system creates a big data predicament that challenges even the most advanced acquisition and storage technologies.

\begin{figure}[!t]
\centering
\includegraphics[width=3.4in]{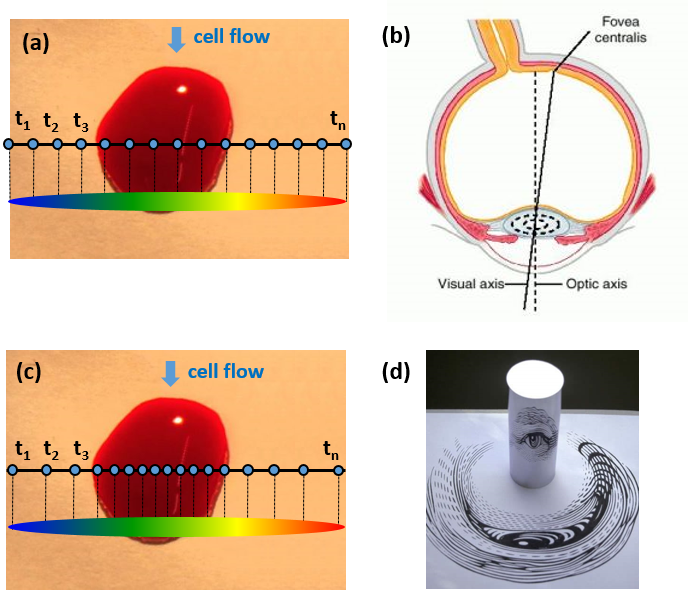}
\caption{An example where the proposed spectrotemporal reshaping leads to efficient sampling. (a) In the STEAM system, spectrum is linearly mapped into time using a dispersive optical fiber with a linear group delay. The temporal waveform is then sampled by a digitizer resulting in uniform spatial sampling. (b) This uniform sampling generates superfluous data by oversampling the sparse peripheral sections of the field of view. (b) The same problem exists in the human vision where high sampling resolution is needed in the central vision while coarse resolution can be tolerated in the peripheral vision. (c) By using nonlinear spectrotemporal mapping via a properly warped group delay the desired nonuniform sampling of the line image is achieved causing efficient allocation of samples to the information-rich regions of the field of view. The reconstruction is a simple dewarping using the inverse of the group delay profile. (d) This process is analogues to image reconstruction in anamorphic art.}
\label{fig:Fovea}
\end{figure}

\figurename~\ref{fig:Fovea} illustrates the big data problem in imaging. The field of view consists of a cell against a background such as a flow channel or a microscope slide. Illumination by an optical pulse that is diffracted into a 1D rainbow maps the 1D space into the optical spectrum (\figurename~\ref{fig:Fovea}a). In the time stretch camera (STEAM), spectrum is linearly mapped into time using a dispersive optical fiber with a linear group delay. The temporal waveform is then sampled by a digitizer resulting in uniform spatial sampling. This uniform sampling generates superfluous data by oversampling the sparse peripheral sections of the field of view, which do not contain significant amounts of information. The same problem exists in the human vision where high sampling resolution is needed in the central vision while coarse resolution can be tolerated in the peripheral vision (\figurename~\ref{fig:Fovea}b). In nature, this is solved by nonuniform photoreceptor density in the retina. The fovea section of the retina has a much higher density of photoreceptors than the rest of the retina and is responsible for the high resolution of central vision necessary for activities where visual detail is of primary importance, such as reading.

We can emulate functionality of the fovea by using nonlinear mapping of spectrum into time via a warped group delay shown in \figurename~\ref{fig:Spectrotemporal_Simulation}d \cite{asghari2014experimental,chen2015optical,jalali2015high,mahjoubfar2015time}. This leads to the desired nonuniform sampling of the image line-scans depicted in \figurename~\ref{fig:Fovea}c causing efficient allocation of samples to the information-rich regions of the field of view. The reconstruction is a simple dewarping using the inverse of the group delay. This operation is analogous to the anamorphic art, where the drawn shape is a stretched and warped version of the true object, yet, the viewer sees the true object upon reflection of the painting from a curved mirror (\figurename~\ref{fig:Fovea}d).

\subsection{Signal-to-Noise Ratio Enhancement}

The warped group delay operations described here modify the spectrotemporal distributions. These modifications can be viewed as changes in the time duration, the bandwidth, and consequently the time-bandwidth product of the envelope as observed in \figurename~\ref{fig:Spectrotemporal_Simulation}. By using proper curvature for the group delay dispersion profile, the envelope time-bandwidth product can be compressed or expanded subject to the sparsity of the signal \cite{jalali2014time,mahjoubfar2015sparsity}.

\begin{figure}[!t]
\centering
\includegraphics[width=2.5in]{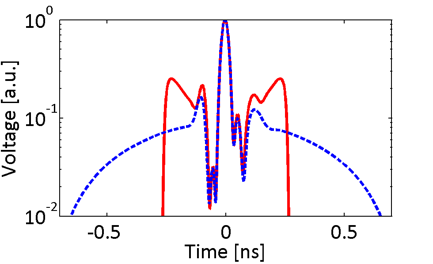}
\caption{Spectrotemporal reshaping leads to the engineering of the signal to noise ratio of an analog signal. Dashed line shows an arbitrary input signal transformed via the $m=2$ mode (linear group delay) and same with the warped group delay in \figurename~\ref{fig:Spectrotemporal_Simulation}d which is synthesized via superposition of this mode with higher order even modes (solid line). By concentrating the energy over a shorter time duration, the warped waveform has a higher instantaneous power.}
\label{fig:SNR}
\end{figure}

Here we show for the first time that for sparse analog signals spectrotemporal reshaping can lead to engineering and improvement of the signal-to-noise ratio (SNR). Note that the SNR enhancement is always relative to the linear dispersion; one can never enhance SNR beyond that of the initial input signal. Also, the improvement is subject to the signal having the proper sparsity such that it can be compressed as discussed previously.  \figurename~\ref{fig:SNR} (dashed line) shows an arbitrary input signal transformed via the $m=2$ mode (linear group delay) and same with the nonlinear group delay in \figurename~\ref{fig:Spectrotemporal_Simulation}d, which is synthesized via superposition of this mode with higher order even modes (solid line). By concentrating the energy over a shorter time duration, the warped waveform has a higher instantaneous power.

Let the average envelope time stretch factor be $M$, and its average bandwidth compression factor be $N$. In the case of a linear group delay ($m=2$) the time-bandwidth product (TBP) is conserved, therefore $N=M$. In the case where it is compressed $N>M$, and in case of expansion, $N<M$. We approximate our system as an adiabatic one, meaning there are no extrinsic energy losses. This is consistent with the fact that optical fibers and fiber based devices that are used to perform the stretch operations have vanishingly low loss. We also assume that the input signal is transform-limited, i.e. it has no chirp. If the input is chirped, the filter group delay profile should be the difference between the desired spectrotemporal chirp and the chirp of the input signal. In the following discussion, signal power and noise refer to the average values over the signal duration.

Upon stretching the waveform's envelope, the average optical power, and the photocurrent, $I$, are reduced by $1/M$. The electrical power is reduced as $1/M^2$. Let us consider the impact of TBP compression or expansion on different noise sources including thermal noise, shot noise, and amplified spontaneous emission noise of optical amplifiers. The analysis below assumes white noise or at least a noise power spectral density that is white over the envelope bandwidth.

\subsubsection{Thermal Noise}

The variance of thermal noise current is
\begin{equation}
\langle i_{th}^2 \rangle = \frac{4 k T}{R} B
\end{equation}
where $R$ is the output resistance and B is the electrical bandwidth. Upon spectrotemporal reshaping, the bandwidth is reduced to $B/N$. Because of the relaxed bandwidth, for a given load capacitance, $C$, the $RC$ time constant can now be increased proportionally. Increasing the load resistance $R$ by $N$ leads to a $1/N^2$ total reduction in thermal noise variance, $\langle i_{th}^2 \rangle$. Therefore the thermal noise limited signal-to-noise ratio (SNR) is scaled by the factor $(1/M^2)/(1/N^2 )=(N/M)^2$. For TBP compression, i.e. when $M<N$, SNR is increased by $(N/M)^2$, whereas it is reduced by the same factor for case of TBP expansion, $M>N$.

\subsubsection{Shot Noise}

The variance of shot noise current is
\begin{equation}
\langle i_{sh}^2 \rangle = 2 I q B
\end{equation}
where $I$ is the average photocurrent, and $q$ is the electron charge. Upon spectrotemporal operation, power is reduced by $M$, and bandwidth is reduced by $N$. Hence the shot noise variance $\langle i_{sh}^2 \rangle$ is reduced by $MN$. Therefore SNR is scaled by the factor $(1/M^2)/(1/M \cdot N)=N/M$. For TBP compression, i.e. when $M<N$, SNR is increased by $N/M$, whereas it is reduced for expansion when $M>N$. 

\subsubsection{Amplified Spontaneous Emission (ASE) Noise} 

In optical systems that employ optical amplification, the SNR is often limited by the amplified spontaneous emission noise and in particular, by the beating of this noise with the signal within the square-law photodetector. The so-called ASE-signal beat noise has a similar behavior to the shot noise and is added to it. The variance of the noise current is then
\begin{equation}
\langle i_{ASE}^2 \rangle = 2 (G I) q B + 4 r_d S_{ASE} (G I) B
\end{equation}
where $r_d=q/h\nu$ is the photodetector responsivity, $h\nu$ is the photon energy, $G$ is the gain of the optical amplifier, $S_{ASE} = n_{sp} h\nu (G-1)$ is the power spectral density of ASE, and $n_{sp} \geq 1$ is the population inversion factor, a parameter that describes how well the requisite population inversion in gain medium has been reached.

As can be seen, the ASE-signal beat noise variance is linearly proportional to the average photocurrent and to the bandwidth, therefore it is similar to the shot noise case. The SNR will be scaled as $N/M$ similar to the case of shot noise. Table \ref{tbl:SNR_improvement} summarizes the above obtained results.

\begin{table}[!t]
\renewcommand{\arraystretch}{1.3}
\caption{Impact of Time-bandwidth Engineering on the Electrical SNR.}
\label{tbl:SNR_improvement}
\centering
\begin{tabular}{|c||c|c|c|}
\hline
& TBP constant & TBP compressed & TBP expanded\\
& $M=N$ & $M<N$ & $M>N$\\
\hhline{|=||=|=|=|}
Thermal noise & No & Increased by & Reduced by\\
limited SNR & change & $(N/M)^2$ & $(N/M)^2$\\
\hline
Shot noise & No & Increased by & Reduced by\\
limited SNR & change & $N/M$ & $N/M$\\
\hline
ASE noise & No & Increased by & Reduced by\\
limited SNR & change & $N/M$ & $N/M$\\
\hline
\multicolumn{4}{p{8.1cm}}{The impact on optical SNR is the same as the electrical SNR listed but to the power $1/2$. TBP: time-bandwidth product; $M$: envelope time stretch ratio; $N$: envelope bandwidth compression ratio. Average compression and expansion over the envelope duration and bandwidth are considered. Note that the SNR enhancement is always relative to the linear dispersion; one can never enhance SNR beyond that of the initial input signal. Results also refer to average SNR over duration and spectrum. Also, the improvement is subject to the signal having the proper sparsity such that it can be compressed as discussed previously. The analysis assumes white noise power spectral density or at least one that is white over the envelope bandwidth.}\\
\end{tabular}
\end{table}

\subsection{Loss in Signal Reconstruction}

We now make an important remark about the ability to reconstruct the signal whose time-bandwidth product has been engineered. For any time-limited pulse such as the pulses analyzed by warped stretch modes, the spectrum is not bandlimited and the signal reconstruction will suffer from the loss of out-of-band spectral components in the acquisition system (photodetector and A/D converter). In far field of dispersion, the bandwidth limitations imposed by the acquisition system can be considered as a frequency-dependent effective bandwidth on the variations of the spectrum (frequency of the spectrum). This effective bandwidth interpretation facilitates the design of group delay profile for a set of desired signals with known spectral characteristics and determines the amount of information lost in the time-bandwidth engineering and reconstruction process.

With the appropriate dispersion design, the spectral features lost in the process of warped stretch become insignificant, and the SNR of the warped stretch is superior to that of the linear dispersion. Note that the SNR enhancement is relative to the linear dispersion. This is not a problem as the application of our method is generally for the ultrashort signals that cannot be measured in real-time without the time stretch. If the desired information is not only in the magnitude of the spectrum but also is in its phase, for both linear and warped group delay dispersions, phase retrieval is necessary, and it increases the reconstruction loss and degrades the SNR in both cases.

\subsection{Spectrotemporal Coding}

As another application, we consider using the dispersion modes to encode information into intra-pulse spectrotemporal characteristics. The group delay dispersion primitives (\figurename~\ref{fig:modes} column 3) induce nonlinear chirp onto a transform limited input pulse. The magnitude and shape of the chirp is dictated by the group delay dispersion curvature of the specific mode. Complex chirp profiles, representing sophisticated codes, can be synthesized using superposition of these primitives. The decoding will consist of operating on the encoded signal with the conjugate of the group delay primitive.  This chirp engineering and coding via group delay dispersion modes will have applications in encryption, secure communication, code division multiple access, and signal processing. 

\subsection{Classification and Feature Detection}

These modes and their corresponding stretch wavelets have unique properties that are used to synthesize operations that are configured for matched response to specific classes of signals with distinct spectrotemporal characteristics. In one application, a bank of dispersive filters, chosen from a library, would be used to probe for presence of specific features of interest. When implemented with optical filters, the analog nature of this processing ensures real-time operation and low power consumption. As discussed below, even-order nonlinear dispersion modes reveal edges and sharp features in the data. Performed in real-time, feature detection can be used to transform the signal into a feature space that is better suited and more efficient for signal classification.

\subsection{Data Compression}

The nonlinear sparse Fourier domain sampling described above may be used for data compression. This works when some frequencies are more important than others. Important frequencies are coded with fine resolution preserving features of spectrum at these frequencies.  On the other hand, less important frequencies are coded with a coarser resolution.  Naturally, some of the finer details of less important frequencies will be lost in the coding.

In the above example, the information of interest is encoded into the amplitude of the spectrum, therefore a simple unwarping of the time-to-spectrum map is sufficient for reconstruction. For a more general case where the information is contained in both the amplitude and phase, reconstruction requires either coherent detection or recovery of phase from amplitude measurements. The input signal is then recovered by simulation of back propagation through the dispersive operator, (\ref{eqn:operator}). Generally known as a phase retrieval method, there are numerous digital algorithms available for recovering the complex amplitude from intensity-only measurements \cite{solli2009optical,asghari2012stereopsis,asghari2012self,dorrer2003complete,walmsley2009characterization,liu1987phase,grilli2001whole,zhang2003whole,chan2008terahertz,jaganathan2012phase}. The dependence of the reconstruction accuracy on the signal-to-noise ratio of the acquired signal has also been analyzed \cite{chan2014reconstruction}. As stated previously, because of noise and limited resolution of analog-to-digital converters, the reconstruction will never be ideal and therefore, this is a lossy compression method.

\subsection{Continuous-time Processing}

In order to apply the proposed transformation for the acquisition and analysis of a continuous-time signal, the signal needs to be segmented into multiple pulse trains using a process called virtual time gating \cite{han2005continuous}. The pulse trains are independently dispersed by linear or warped time stretch systems in parallel, and the acquired signals are digitally concatenated to reveal the spectral features of the continuous-time signal as it varies with time. The time duration of the gate and the shape of the dispersion profile can be reconfigured for application to nonstationary signals. 

\subsection{Digital Implementation and Event Detection for Real-time Analytics}
\label{scn:digital_implementation}
The signal transformations described here can be implemented numerically and applied to digital data.  The one-dimensional temporal data can be generalized to 2D images as well as N-dimensional data. In image processing applications, optical diffraction takes the place of temporal dispersion of the envelope, and the mathematics will be a 2D discrete version of the 1D continuous case presented above (image spectrum is in baseband, therefore a more accurate analogy is to baseband carrier-less diffraction). The result will be a physics-inspired digital signal processing technique that achieves multi-dimensional warped Fourier domain sampling, by emulating propagation of electromagnetic waves through a diffractive medium with an engineered dielectric function. As was done with temporal dispersion here, diffraction can be modeled by an all-pass phase filter with specific frequency dependencies characterized by the constituent dispersion modes. Numerical implementation also makes it possible to use arbitrary dispersion modes and sophisticated spectrotemporal basis sets, such as pseudorandom sets, that would be difficult to achieve in a physical system.

The magnitude of the transformed data leads to nonuniform Fourier transformation. The phase reveals sharp transitions and offers a categorically new method for edge detection in images \cite{asghari2014physics,asghari2015edge}. Edge detection is used for identifying patterns in digital images where brightness or color changes abruptly. Applying an edge detection algorithm to an image can be used for object detection and classification. It also reduces the digital file size while preserving important information, albeit data compression is not the main objective in edge detection.  

\begin{figure}[!t]
\centering
\includegraphics[width=2.5in]{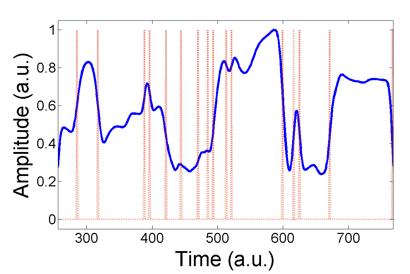}
\caption{Detection of transients and edges in a waveform. Here the temporal phase of the transformed signal contains information about fast transitions and reveals edges and anomalies. An arbitrary temporal waveform as well as the phase of the transformed signal clearly showing the ability to identify transitions in the waveform. This can be used for edge triggering, pattern recognition, and event detection.}
\label{fig:Edge}
\end{figure}

The same edge function can also be performed on temporal waveforms in analog domain to reveal transitions and anomalies in real-time as shown in \figurename~\ref{fig:Edge}. This is performed by operating on the signal with even-order dispersive modes (see \figurename~\ref{fig:modes}). Applications include edge triggering, pattern recognition, and event detection. Similar to its use in digital image processing \cite{asghari2014physics,asghari2015edge}, analog edge detection can be used for compression by reducing the data set while preserving important information. 

\section{Conclusion}
In summary, we have introduced the concepts of photonic hardware accelerators and one rendition of them in terms of real-time analog information engineering units that are based on nonlinear dispersion modes. These modes are basic building blocks of complex spectrotemporal transformations that cause large amounts of nonlinear frequency dependent phase (warped chirp) using curved group delay dispersion. While this requirement is difficult to realize in analog electronics over broad bandwidth and with low loss, it is readily realized in optics using dispersive devices that provide large group delay dispersion, low loss, and reconfigurability. The requirements are also readily met in the digital (numerical) implementation. 

The inherent nonlinearity (square-law) in the optical to electrical conversion is an intrinsic part of the proposed signal transformations. In one class of applications, the operation leads to warped frequency-to-time mapping. The warp profile is matched to the signal's spectrum sparsity such that, when followed by a uniform sampler, the reshaping of the signal results in nonuniform Fourier domain sampling with more samples allocated to the dense portions of the spectrum and fewer to the sparse regions. In another application, the operations performed using optical dispersive elements lead to compression in time-bandwidth product of an analog signal with accompanying improvement in signal-to-noise ratio in detection and analog-to-digital conversion. Yet in a different implementation the nonlinear dispersion modes can be used for spectrotemporal coding, feature detection, and classification. We anticipate that further research on concepts of nonlinear dispersion modes and photonic hardware accelerator introduced here will lead to additional applications.

\section*{Acknowledgment}

This work was partially supported by the Office of Naval Research (ONR) Multidisciplinary University Research Initiatives (MURI) Program on optical computing. We are grateful to Daniel Solli at University of California, Los Angeles for valuable discussions. 

% Can use something like this to put references on a page
% by themselves when using endfloat and the captionsoff option.
\ifCLASSOPTIONcaptionsoff
  \newpage
\fi

\bibliographystyle{IEEEtran}
% argument is your BibTeX string definitions and bibliography database(s)
\bibliography{IEEEabrv,TheProceedings}

% Generated by IEEEtran.bst, version: 1.13 (2008/09/30)
\begin{thebibliography}{10}
\providecommand{\url}[1]{#1}
\csname url@samestyle\endcsname
\providecommand{\newblock}{\relax}
\providecommand{\bibinfo}[2]{#2}
\providecommand{\BIBentrySTDinterwordspacing}{\spaceskip=0pt\relax}
\providecommand{\BIBentryALTinterwordstretchfactor}{4}
\providecommand{\BIBentryALTinterwordspacing}{\spaceskip=\fontdimen2\font plus
\BIBentryALTinterwordstretchfactor\fontdimen3\font minus
  \fontdimen4\font\relax}
\providecommand{\BIBforeignlanguage}[2]{{%
\expandafter\ifx\csname l@#1\endcsname\relax
\typeout{** WARNING: IEEEtran.bst: No hyphenation pattern has been}%
\typeout{** loaded for the language `#1'. Using the pattern for}%
\typeout{** the default language instead.}%
\else
\language=\csname l@#1\endcsname
\fi
#2}}
\providecommand{\BIBdecl}{\relax}
\BIBdecl

\bibitem{owens2007survey}
J.~D. Owens, D.~Luebke, N.~Govindaraju, M.~Harris, J.~Kr{\"u}ger, A.~E. Lefohn,
  and T.~J. Purcell, ``A survey of general-purpose computation on graphics
  hardware,'' in \emph{Computer graphics forum}, vol.~26, no.~1.\hskip 1em plus
  0.5em minus 0.4em\relax Wiley Online Library, 2007, pp. 80--113.

\bibitem{manavski2007cuda}
S.~A. Manavski, ``Cuda compatible gpu as an efficient hardware accelerator for
  aes cryptography,'' in \emph{Signal Processing and Communications, 2007.
  ICSPC 2007. IEEE International Conference on}.\hskip 1em plus 0.5em minus
  0.4em\relax IEEE, 2007, pp. 65--68.

\bibitem{porrmann2002implementation}
M.~Porrmann, U.~Witkowski, H.~Kalte, and U.~R{\"u}ckert, ``Implementation of
  artificial neural networks on a reconfigurable hardware accelerator,'' in
  \emph{16th Euromicro Conference on Parallel, Distributed and Network-Based
  Processing (PDP 2008)}.\hskip 1em plus 0.5em minus 0.4em\relax IEEE Computer
  Society, 2002, pp. 0243--0243.

\bibitem{bosworth2013high}
B.~T. Bosworth and M.~A. Foster, ``High-speed ultrawideband photonically
  enabled compressed sensing of sparse radio frequency signals,'' \emph{Optics
  letters}, vol.~38, no.~22, pp. 4892--4895, 2013.

\bibitem{valley2012compressive}
G.~C. Valley, G.~A. Sefler, and T.~J. Shaw, ``Compressive sensing of sparse
  radio frequency signals using optical mixing,'' \emph{Optics letters},
  vol.~37, no.~22, pp. 4675--4677, 2012.

\bibitem{le1991mpeg}
D.~Le~Gall, ``Mpeg: A video compression standard for multimedia applications,''
  \emph{Communications of the ACM}, vol.~34, no.~4, pp. 46--58, 1991.

\bibitem{richardson2004h}
I.~E. Richardson, \emph{H. 264 and MPEG-4 video compression: video coding for
  next-generation multimedia}.\hskip 1em plus 0.5em minus 0.4em\relax John
  Wiley \& Sons, 2004.

\bibitem{coppinger1998time}
F.~Coppinger, A.~Bhushan, and B.~Jalali, ``Time magnification of electrical
  signals using chirped optical pulses,'' \emph{Electronics Letters}, vol.~34,
  no.~4, pp. 399--400, 1998.

\bibitem{kelkar1999time}
P.~Kelkar, F.~Coppinger, A.~Bhushan, and B.~Jalali, ``Time-domain optical
  sensing,'' \emph{Electronics Letters}, vol.~35, no.~19, pp. 1661--1662, 1999.

\bibitem{solli2007optical}
D.~Solli, C.~Ropers, P.~Koonath, and B.~Jalali, ``Optical rogue waves,''
  \emph{Nature}, vol. 450, no. 7172, pp. 1054--1057, 2007.

\bibitem{Ng2014demonstration}
W.~Ng, T.~Rockwood, and A.~Reamon, ``Demonstration of channel-stitched photonic
  time stretch analog-to-digital converter with enob ≥ 8 for a 10 ghz signal
  bandwidth,'' in \emph{GOMACTech}.\hskip 1em plus 0.5em minus 0.4em\relax US
  Department of Defense, 2014, p. 26.2.

\bibitem{goda2009serial}
K.~Goda, K.~Tsia, and B.~Jalali, ``Serial time-encoded amplified imaging for
  real-time observation of fast dynamic phenomena,'' \emph{Nature}, vol. 458,
  no. 7242, pp. 1145--1149, 2009.

\bibitem{qian2009real}
F.~Qian, Q.~Song, E.-k. Tien, S.~K. Kalyoncu, and O.~Boyraz, ``Real-time
  optical imaging and tracking of micron-sized particles,'' \emph{Optics
  Communications}, vol. 282, no.~24, pp. 4672--4675, 2009.

\bibitem{zhang2011serial}
C.~Zhang, Y.~Qiu, R.~Zhu, K.~K. Wong, and K.~K. Tsia, ``Serial time-encoded
  amplified microscopy (steam) based on a stabilized picosecond supercontinuum
  source,'' \emph{Optics express}, vol.~19, no.~17, pp. 15\,810--15\,816, 2011.

\bibitem{fard2011nomarski}
A.~M. Fard, A.~Mahjoubfar, K.~Goda, D.~R. Gossett, D.~Di~Carlo, and B.~Jalali,
  ``Nomarski serial time-encoded amplified microscopy for high-speed
  contrast-enhanced imaging of transparent media,'' \emph{Biomedical optics
  express}, vol.~2, no.~12, pp. 3387--3392, 2011.

\bibitem{wong2012optical}
T.~T. Wong, A.~K. Lau, K.~K. Wong, and K.~K. Tsia, ``Optical time-stretch
  confocal microscopy at 1 $\mu$m,'' \emph{Optics letters}, vol.~37, no.~16,
  pp. 3330--3332, 2012.

\bibitem{kalyoncu2014fast}
S.~K. Kalyoncu, R.~Torun, Y.~Huang, Q.~Zhao, and O.~Boyraz, ``Fast dispersive
  laser scanner by using digital micro mirror arrays,'' \emph{Journal of Micro
  and Nano-Manufacturing}, vol.~2, no.~2, p. 021004, 2014.

\bibitem{goda2012high}
K.~Goda, A.~Ayazi, D.~R. Gossett, J.~Sadasivam, C.~K. Lonappan, E.~Sollier,
  A.~M. Fard, S.~C. Hur, J.~Adam, C.~Murray \emph{et~al.}, ``High-throughput
  single-microparticle imaging flow analyzer,'' \emph{Proceedings of the
  National Academy of Sciences}, vol. 109, no.~29, pp. 11\,630--11\,635, 2012.

\bibitem{mahjoubfar2013label}
A.~Mahjoubfar, C.~Chen, K.~R. Niazi, S.~Rabizadeh, and B.~Jalali, ``Label-free
  high-throughput cell screening in flow,'' \emph{Biomedical optics express},
  vol.~4, no.~9, pp. 1618--1625, 2013.

\bibitem{chen2014hyper}
C.~Chen, A.~Mahjoubfar, A.~Huang, K.~Niazi, S.~Rabizadeh, and B.~Jalali,
  ``Hyper-dimensional analysis for label-free high-throughput imaging flow
  cytometry,'' in \emph{CLEO: Applications and Technology}.\hskip 1em plus
  0.5em minus 0.4em\relax Optical Society of America, 2014, pp. AW3L--2.

\bibitem{mahjoubfar2014label}
A.~Mahjoubfar, C.~Chen, K.~Niazi, S.~Rabizadeh, and B.~Jalali, ``Label-free
  high-throughput imaging flow cytometry,'' in \emph{SPIE LASE}.\hskip 1em plus
  0.5em minus 0.4em\relax International Society for Optics and Photonics, 2014,
  pp. 89\,720F--89\,720F.

\bibitem{mahjoubfar2011high}
A.~Mahjoubfar, K.~Goda, A.~Ayazi, A.~Fard, S.~H. Kim, and B.~Jalali,
  ``High-speed nanometer-resolved imaging vibrometer and velocimeter,''
  \emph{Applied Physics Letters}, vol.~98, no.~10, p. 101107, 2011.

\bibitem{yazaki2014ultrafast}
A.~Yazaki, C.~Kim, J.~Chan, A.~Mahjoubfar, K.~Goda, M.~Watanabe, and B.~Jalali,
  ``Ultrafast dark-field surface inspection with hybrid-dispersion laser
  scanning,'' \emph{Applied Physics Letters}, vol. 104, no.~25, p. 251106,
  2014.

\bibitem{goda2012hybrid}
K.~Goda, A.~Mahjoubfar, C.~Wang, A.~Fard, J.~Adam, D.~R. Gossett, A.~Ayazi,
  E.~Sollier, O.~Malik, E.~Chen \emph{et~al.}, ``Hybrid dispersion laser
  scanner,'' \emph{Scientific reports}, vol.~2, 2012.

\bibitem{mahjoubfar20133d}
A.~Mahjoubfar, K.~Goda, C.~Wang, A.~Fard, J.~Adam, D.~Gossett, A.~Ayazi,
  E.~Sollier, O.~Malik, E.~Chen \emph{et~al.}, ``3d ultrafast laser scanner,''
  in \emph{SPIE LASE}.\hskip 1em plus 0.5em minus 0.4em\relax International
  Society for Optics and Photonics, 2013, pp. 86\,110N--86\,110N.

\bibitem{goda2013dispersive}
K.~Goda and B.~Jalali, ``Dispersive fourier transformation for fast continuous
  single-shot measurements,'' \emph{Nature Photonics}, vol.~7, no.~2, pp.
  102--112, 2013.

\bibitem{goda2009theory}
K.~Goda, D.~R. Solli, K.~K. Tsia, and B.~Jalali, ``Theory of amplified
  dispersive fourier transformation,'' \emph{Physical Review A}, vol.~80,
  no.~4, p. 043821, 2009.

\bibitem{agrawal2007nonlinear}
G.~P. Agrawal, \emph{Nonlinear fiber optics}.\hskip 1em plus 0.5em minus
  0.4em\relax Academic press, 2007.

\bibitem{chou2007femtosecond}
J.~Chou, O.~Boyraz, D.~Solli, and B.~Jalali, ``Femtosecond real-time
  single-shot digitizer,'' \emph{Applied Physics Letters}, vol.~91, no.~16, pp.
  161\,105--161\,105, 2007.

\bibitem{mahjoubfar2013optically}
A.~Mahjoubfar, K.~Goda, G.~Betts, and B.~Jalali, ``Optically amplified
  detection for biomedical sensing and imaging,'' \emph{JOSA A}, vol.~30,
  no.~10, pp. 2124--2132, 2013.

\bibitem{goda2009demonstration}
K.~Goda, A.~Mahjoubfar, and B.~Jalali, ``Demonstration of raman gain at 800 nm
  in single-mode fiber and its potential application to biological sensing and
  imaging,'' \emph{Applied Physics Letters}, vol.~95, no.~25, p. 251101, 2009.

\bibitem{hill1997fiber}
K.~O. Hill and G.~Meltz, ``Fiber bragg grating technology fundamentals and
  overview,'' \emph{Journal of lightwave technology}, vol.~15, no.~8, pp.
  1263--1276, 1997.

\bibitem{conway2008phase}
J.~A. Conway, G.~A. Sefler, J.~T. Chou, and G.~C. Valley, ``Phase ripple
  correction: theory and application,'' \emph{Optics letters}, vol.~33, no.~10,
  pp. 1108--1110, 2008.

\bibitem{diebold2011giant}
E.~D. Diebold, N.~K. Hon, Z.~Tan, J.~Chou, T.~Sienicki, C.~Wang, and B.~Jalali,
  ``Giant tunable optical dispersion using chromo-modal excitation of a
  multimode waveguide,'' \emph{Optics express}, vol.~19, no.~24, pp.
  23\,809--23\,817, 2011.

\bibitem{klein2005reconfigurable}
E.~J. Klein, D.~H. Geuzebroek, H.~Kelderman, G.~Sengo, N.~Baker, and
  A.~Driessen, ``Reconfigurable optical add-drop multiplexer using microring
  resonators,'' \emph{Photonics Technology Letters, IEEE}, vol.~17, no.~11, pp.
  2358--2360, 2005.

\bibitem{riza1999reconfigurable}
N.~A. Riza and S.~Yuan, ``Reconfigurable wavelength add-drop filtering based on
  a banyan network topology and ferroelectric liquid crystal fiber-optic
  switches,'' \emph{Journal of lightwave technology}, vol.~17, no.~9, p. 1575,
  1999.

\bibitem{park2015dispersion}
H.~Park, M.~Ashgari, and B.~Jalali, ``Dispersion engineering employing curved
  space mapping and chromo-modal excitation,'' in \emph{CLEO: Applications and
  Technology}.\hskip 1em plus 0.5em minus 0.4em\relax Optical Society of
  America, 2015, p. Accepted for publication.

\bibitem{xing2014serial}
F.~Xing, H.~Chen, C.~Lei, Z.~Weng, M.~Chen, S.~Yang, and S.~Xie, ``Serial
  wavelength division 1 ghz line-scan microscopic imaging,'' \emph{Photonics
  Research}, vol.~2, no.~4, pp. B31--B34, 2014.

\bibitem{xing20152}
F.~Xing, H.~Chen, C.~Lei, M.~Chen, S.~Yang, and S.~Xie, ``A 2-ghz
  discrete-spectrum waveband-division microscopic imaging system,''
  \emph{Optics Communications}, vol. 338, pp. 22--26, 2015.

\bibitem{asghari2014experimental}
M.~H. Asghari and B.~Jalali, ``Experimental demonstration of optical real-time
  data compression,'' \emph{Applied Physics Letters}, vol. 104, no.~11, p.
  111101, 2014.

\bibitem{chen2015optical}
C.~L. Chen, A.~Mahjoubfar, and B.~Jalali, ``Optical data compression in time
  stretch imaging,'' \emph{PLoS ONE}, vol.~10, no. Accepted for publication, p.
  Accepted for publication, 2015.

\bibitem{jalali2015high}
B.~Jalali, A.~Mahjoubfar, and C.~L. Chen, ``High-throughput biological cell
  classification featuring real-time optical data compression,'' in
  \emph{Conference on Information Sciences and Systems, Baltimore, MD}.\hskip
  1em plus 0.5em minus 0.4em\relax IEEE, 2015, p. Accepted for publication.

\bibitem{mahjoubfar2015time}
A.~Mahjoubfar, C.~L. Chen, and B.~Jalali, ``Time stretch imaging with optical
  data compression for label-free biological cell classification,'' in
  \emph{Opto Electronics and Communications Conference, Shanghai, China}.\hskip
  1em plus 0.5em minus 0.4em\relax IEEE, 2015, p. Accepted for publication.

\bibitem{jalali2014time}
B.~Jalali, J.~Chan, and M.~H. Asghari, ``Time--bandwidth engineering,''
  \emph{Optica}, vol.~1, no.~1, pp. 23--31, 2014.

\bibitem{mahjoubfar2015sparsity}
A.~Mahjoubfar, J.~Chan, M.~H. Asghari, and B.~Jalali, ``Sparsity and
  self-adaptivity in anamorphic stretch transform,'' in \emph{Conference on
  Information Sciences and Systems, Baltimore, MD}.\hskip 1em plus 0.5em minus
  0.4em\relax IEEE, 2015, p. Accepted for publication.

\bibitem{solli2009optical}
D.~Solli, S.~Gupta, and B.~Jalali, ``Optical phase recovery in the dispersive
  fourier transform,'' \emph{Applied Physics Letters}, vol.~95, no.~23, p.
  231108, 2009.

\bibitem{asghari2012stereopsis}
M.~H. Asghari and B.~Jalali, ``Stereopsis-inspired time-stretched amplified
  real-time spectrometer (stars),'' \emph{Photonics Journal, IEEE}, vol.~4,
  no.~5, pp. 1693--1701, 2012.

\bibitem{asghari2012self}
M.~H. Asghari and J.~Aza{\~n}a, ``Self-referenced temporal phase reconstruction
  from intensity measurements using causality arguments in linear optical
  filters,'' \emph{Optics letters}, vol.~37, no.~17, pp. 3582--3584, 2012.

\bibitem{dorrer2003complete}
C.~Dorrer and I.~Kang, ``Complete temporal characterization of short optical
  pulses by simplified chronocyclic tomography,'' \emph{Optics letters},
  vol.~28, no.~16, pp. 1481--1483, 2003.

\bibitem{walmsley2009characterization}
I.~A. Walmsley and C.~Dorrer, ``Characterization of ultrashort electromagnetic
  pulses,'' \emph{Advances in Optics and Photonics}, vol.~1, no.~2, pp.
  308--437, 2009.

\bibitem{liu1987phase}
G.~Liu and P.~Scott, ``Phase retrieval and twin-image elimination for in-line
  fresnel holograms,'' \emph{JOSA A}, vol.~4, no.~1, pp. 159--165, 1987.

\bibitem{grilli2001whole}
S.~Grilli, P.~Ferraro, S.~De~Nicola, A.~Finizio, G.~Pierattini, and R.~Meucci,
  ``Whole optical wavefields reconstruction by digital holography,''
  \emph{Optics Express}, vol.~9, no.~6, pp. 294--302, 2001.

\bibitem{zhang2003whole}
Y.~Zhang, G.~Pedrini, W.~Osten, and H.~Tiziani, ``Whole optical wave field
  reconstruction from double or multi in-line holograms by phase retrieval
  algorithm,'' \emph{Optics Express}, vol.~11, no.~24, pp. 3234--3241, 2003.

\bibitem{chan2008terahertz}
W.~L. Chan, M.~L. Moravec, R.~G. Baraniuk, and D.~M. Mittleman, ``Terahertz
  imaging with compressed sensing and phase retrieval,'' \emph{Optics letters},
  vol.~33, no.~9, pp. 974--976, 2008.

\bibitem{jaganathan2012phase}
K.~Jaganathan, S.~Oymak, and B.~Hassibi, ``Phase retrieval for sparse signals
  using rank minimization,'' in \emph{Acoustics, Speech and Signal Processing
  (ICASSP), 2012 IEEE International Conference on}.\hskip 1em plus 0.5em minus
  0.4em\relax IEEE, 2012, pp. 3449--3452.

\bibitem{chan2014reconstruction}
J.~Chan, A.~Mahjoubfar, M.~H. Asghari, and B.~Jalali, ``Reconstruction in
  time-bandwidth compression systems,'' \emph{Applied Physics Letters}, vol.
  105, no.~22, p. 221105, 2014.

\bibitem{han2005continuous}
Y.~Han and B.~Jalali, ``Continuous-time time-stretched analog-to-digital
  converter array implemented using virtual time gating,'' \emph{Circuits and
  Systems I: Regular Papers, IEEE Transactions on}, vol.~52, no.~8, pp.
  1502--1507, 2005.

\bibitem{asghari2014physics}
M.~H. Asghari and B.~Jalali, ``Physics-inspired image edge detection,'' in
  \emph{Signal and Information Processing (GlobalSIP), 2014 IEEE Global
  Conference on}.\hskip 1em plus 0.5em minus 0.4em\relax IEEE, 2014, pp.
  293--296.

\bibitem{asghari2015edge}
------, ``Edge detection in digital images using dispersive phase stretch
  transform,'' \emph{International Journal of Biomedical Imaging}, vol. 2015,
  p. 687819, 2015.

\end{thebibliography}

\begin{IEEEbiography}[{\includegraphics[width=1in,height=1.25in,clip,keepaspectratio]{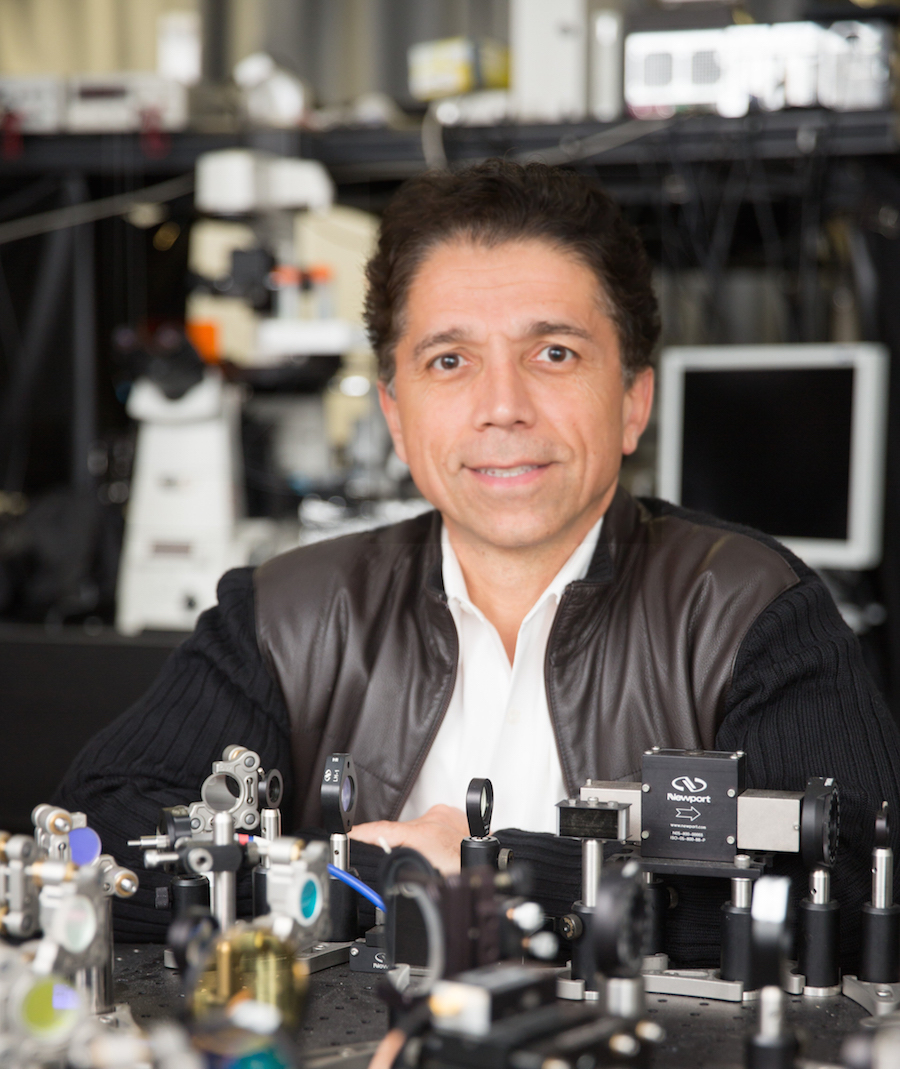}}]{Bahram Jalali}
is the Northrop-Grumman Endowed Chair in Optoelectronics and Professor of Electrical Engineering at UCLA with joint appointments in the Biomedical Engineering and in the UCLA David Geffen School of Medicine. He is a Fellow of IEEE, the Optical Society of America (OSA), the American Physical Society (APS) and SPIE. He is the recipient of the R. W. Wood Prize from Optical Society of America for the invention and demonstration of the first Silicon Laser, and the Aron Kressel Award of the IEEE Photonics Society, the IET Achievement Medal, and the Distinguished Engineering Achievement Award from the Engineers Council.
\end{IEEEbiography}

\begin{IEEEbiography}[{\includegraphics[width=1in,height=1.25in,clip,keepaspectratio]{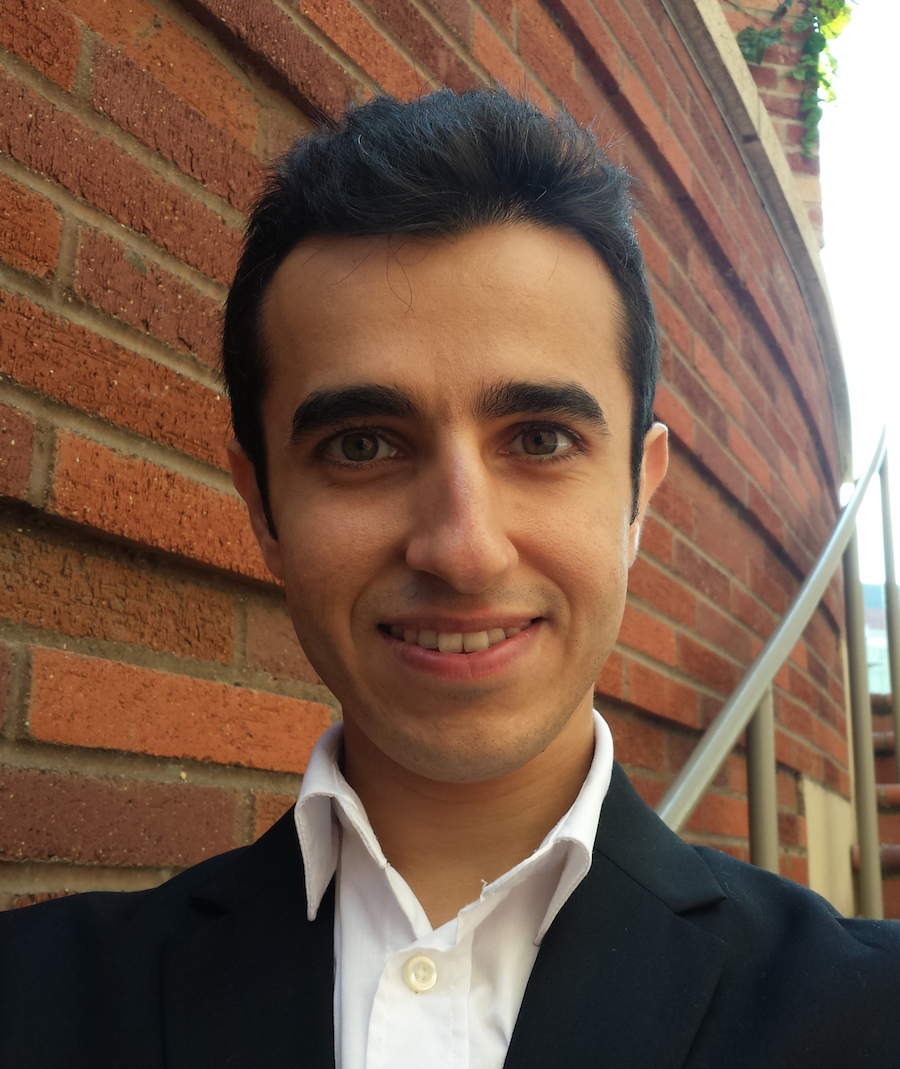}}]{Ata Mahjoubfar}
received his Ph.D. degree in electrical engineering from University of California, Los Angeles (UCLA) in 2014, where he is currently a postdoctoral scholar. He was the cofounder of OSA/SPIE student chapter at UCLA and its president in 2012. He is the author of more than twenty peer-reviewed publications. His research interests include ultrafast data analytics, image and signal processing, machine vision and learning, imaging and display technologies, and biomedical engineering.
\end{IEEEbiography}

% insert where needed to balance the two columns on the last page with
% biographies
%\newpage

% You can push biographies down or up by placing
% a \vfill before or after them. The appropriate
% use of \vfill depends on what kind of text is
% on the last page and whether or not the columns
% are being equalized.

%\vfill

% Can be used to pull up biographies so that the bottom of the last one
% is flush with the other column.
%\enlargethispage{-3in}

\end{document}